\documentclass[12pt]{iopart}
\usepackage{fleqn,epsfig}
\usepackage{graphicx}
\usepackage{slashed}

\def\be{\begin{equation}}
\def\ee{\end{equation}}
\def\br{\begin{eqnarray}}
\def\er{\end{eqnarray}}
\def\bc{\begin{center}}
\def\ec{\end{center}}

\def\L {{{\cal L}}}

\def\N3{{1\over (2\pi)^3}}
\def\rf#1{{(\ref{#1})}}

\def\ket#1{|#1 \rangle}
\def\bra#1{\langle #1|}

\def\g5{\gamma_5}

\def\endauthors{}
\def\authors#1\endauthors{#1}

\def\g {{\gamma}}

\def\endauthors{}
\def\authors#1\endauthors{#1}

\def\be{\begin{equation}}
\def\ee{\end{equation}}
\def\br{\begin{eqnarray}}
\def\er{\end{eqnarray}}
\def\brn{\begin{eqnarray*}}
\def\ern{\end{eqnarray*}}
\def\rf#1{{(\ref{#1})}}
\def\ket#1{|#1 \rangle}
\def\bra#1{\langle #1|}

\def\lbar{\mbox{$\lambda$\kern-0,450em \vrule width0,35em height1,252ex
depth-1,21ex \kern0,051em}}
\def\dbar{\mbox{d\kern-0,347em \vrule width0,3em height1,252ex depth-1,21ex
\kern0,051em}}
\def\Dbar{\mbox{D\kern-0,735em \vrule width0,3em height0,86ex depth-0,81ex
\kern0,40em}}

\def\xb {{\bf x}}

\def\g {{\gamma}}

\def\L {{{\cal L}}}

\def\N {{{\cal N}}}

\def\ba#1{\begin{array}{#1}}
\def\ea{\end{array}}

\def\bc{binomial coefficients }

\def\bin#1#2{\left(\negthinspace\begin{array}{c}#1\\#2\end{array}\right)}

\def\be{\begin{equation}}
\def\ee{\end{equation}}
\def\br{\begin{eqnarray}}
\def\er{\end{eqnarray}}
\def\brn{\begin{eqnarray*}}
\def\ern{\end{eqnarray*}}
\def\bit{\begin{itemize}}
\def\eit{\end{itemize}}
\def\bnu{\begin{enumerate}}
\def\enu{\end{enumerate}}

\def\={{\simeq}}
\def\rf#1{{(\ref{#1})}}

\def\nn{\nonumber }

\def\ket#1{|#1 \rangle}
\def\bra#1{\langle #1|}

\def\2q{{{\{}2{\}}_q}}
\def\3q{{{\{}3{\}}_q}}

\def\xb {{\bf x}}

\def\mbga{\mbox{\boldmath$\gamma$}}
\def\mbna{\mbox{\boldmath$\nabla$}}

\begin{document}
\title{Inconsistency of the interactions between pseudoscalar, spinor and Rarita-Schwinger fields}
\author{D. Badagnani$^1$, A. Mariano$^{1,2}$ and C. Barbero$^{1,2}$}



\begin{abstract}

We perform the Dirac quantization of RS fields interacting with a spinor and the first derivative of a pseudoscalar field. We achieve
the calculations for two forms of this interaction: first we review the conventional coupling of lowest   derivative order, reproducing
the well known inconsistencies in its anticommutator algebra. Then, we perform the analysis on the next order term popularly  known as
``spin 3/2 gauge invariant interaction'', which is claimed to be free of these inconsistencies.
Nevertheless we find that the direct application of the Dirac formalism
leads to inconsistencies in complete analogy to the previous case.
This is of high relevance in the  particle phenomenology field, where these interactions are used to interpret experimental data
involving $\Delta (1232)$ resonances.

\end{abstract}

\pacs{11.10.-z	, 11.10.Ef.}

\maketitle

\section{INTRODUCTION }

The problem of setting consistent interactions for higher spin fields has been a much debated subject for
several decades, both in the Quantum Field Theory (QFT) and particle phenomenology communities. From a
theoretical point of view, there are plenty of problems for quantizing such a theory when background fields
are considered: breaking of Lorentz invariance, superluminar propagation, indefinite anticommutators of
mutually conjugate fields and so on. Nevertheless, there are not such problems in perturbation theory and
absence of background\cite{Weinberg_RS}, so its relevance in phenomenology is debatable.

The difficulties are tightly related to the occurrence of constraints. Since vector-spinor fields contains
both a spin 3/2 sector and two spin 1/2 ones, the correct description of spin 3/2 degrees of freedom requires
projection onto the first sector. Nevertheless, the complete space is needed to invert the propagator, so virtual spin 1/2 states
do also  propagate. When quantizing this theory
second class constraints arise, which amount to projecting out the Hilbert space sectors corresponding to the  lower spin.
 But interactions, in general, change the constraints quite drastically making field anticommutators dependent on the dynamics\cite{JS, Hagen}. That  is why one talks of ``quantizing'' the interaction.

The problem was first described in \cite{JS}, for the RS field minimally coupled to the electromagnetic (EM) field. Then, in
\cite{Nath} it was claimed that a linear coupling to a spinor and the derivative of a scalar ($g_{00}=1,g_{ij}=-\delta_{ij}$ and isospin omitted)
\begin{equation}
{\cal L}_{NEK} = g
    \bar{\Psi}^\mu \left( g_{\mu \nu} +
    \left[\frac{1}{2}(1+4Z)A+Z\right] \gamma_\mu \gamma_\nu  \right) \psi \partial_\nu \phi + c.c.
\end{equation}
(where the value $Z=\frac{1}{2}$ was chosen from field consistency theoretical arguments) would be free of such problems, but it was shown by Hagen \cite{Hagen} that
this is not the case. Later, it was shown \cite{Singh} that in the presence of scalar gradients noncausal propagation arises.
The source of the problem was made clear in \cite{AKT}: they have shown that any coupling
leading to linear constraints on the fermion degrees of freedom leads to indefinite anticommutators due to the existence of negative parameters (which they called ``negative masses'') in the kinetic terms of the Lagrangian, which are always present for the RS fields. This may be  construed as a consequence of the reintroduction of the spin 1/2 sector, to which correspond such parameters, which are projected out in the free theory.

More recently, \cite{PASCA} proposed a new interaction which is derivative in the RS field. The most general such
term preserving chiral symmetry can be written as ($\epsilon^{0123}=1,\gamma_5=i\gamma^0\gamma^1\gamma^2\gamma^3$)
\begin{equation}
{\cal L}_{P} = g \bar{\Psi}^\mu \left( g_{\mu \sigma} +
                  \left[\frac{1}{2}(1+4Z)A+Z\right] \gamma_\mu \gamma_\sigma  \right)
                \epsilon^{\sigma \nu\lambda \rho } \gamma_5 \gamma_\lambda (\partial_\rho \psi) (\partial_\nu \phi) + c.c.
\label{LPASCA}
\end{equation}
If the off-shell parameter $Z$ is set to $Z=-\frac{1}{2}$ this is the interaction proposed in \cite{PASCA}.
We show in Appendix A that this value is indeed needed for consistency, so we will not consider other values.
Recall that in the RS formalism the parameter $A$ is unobservable and thus arbitrary. The interaction (\ref{LPASCA})
for $Z=1/2$ is just the one proposed in \cite{PASCA} as rewritten in \cite{Lenske} more generally to restore its $A$ dependence.
This interaction with $Z$ chosen as above has the property of projecting out
spin 1/2 virtual state of the propagator in elastic amplitudes at tree level, but the off-shell sector of spin 1/2 is potentially
present, and manifests itself in radiative amplitudes, as stated below.
It has been argued in \cite{PASCA} that such interactions are free of the above mentioned problems, although in its conclusions
section this assertion is given the status of a conjecture.
This proposal became
quite popular for the description of $\Delta$ resonances.
Nonetheless, we showed recently that in the presence of the EM coupling  (unavoidable in this context, since the $\Delta$ is charged)
the consistency problem remains even for this new proposed interaction:
when the EM interaction is introduced, renormalization considerations force to
reintroduce an interaction of the form ${\cal L}_{NEK}$ \cite{Nos_background}.
The interaction obtained from ${\cal L}_{P}$ does not  eliminate spin 1/2 virtual states
in all circumstances,  radiative processes for instance exhibit a spin 1/2 ''background''. Also, the new interaction is not superior even phenomenologically,
since a background compatible with exchange of virtual spin 1/2 is indeed observed, and the use of ${\cal L}_{NEK}$ vertexes is found to fit better the data than the ${\cal L}_P$ ones
\cite{Nos}. Finally, this new interaction presents also problems with the coexistence with the electromagnetic  gauge invariance \cite{Nos_background}.

However, it remains the interesting theoretical possibility that, in absence of electromagnetic interactions,
${\cal L}_P$ be indeed consistent.
Nevertheless, there are reasons to strongly suspect that it is not the case. Indeed, in spite of being inspired in a
''gauge invariance'' of the kinetic $\Delta$ term,
${\cal L}_P$ can be obtained by simply invoking  the next order interaction (in derivatives) to ${\cal L}_{NEK}$,
which was not considered in \cite{Nath}.
We would thus expect a somewhat more involved but otherwise analogous constraint structure. In fact, the theory exhibit the same
linear constraints in
fermionic degrees of freedom described in \cite{AKT}, so the same positivity issue should have to arise.
In \cite{PASCA} there is some argumentation in favor
of the consistency of ${\cal L}_P$, but  while  the constraint arguments showing  ${\cal L}_{NEK}$ inconsistency are developed with
certain detail, the same analysis was not performed for ${\cal L}_P$. Instead, a ''St\"uckelberg parameter''
is introduced in order to render
the massive theory ''spin 3/2 gauge invariant''. Nevertheless,  that's not the right procedure since  a St\"uckelberg variable is not a
parameter but a dynamical field \cite{Stuec} (see Appendix C). So, a complete constraints analysis treating both
${\cal L}_{NEK}$ and ${\cal L}_P$ on the same footing is desirable.

The paper is organized as follows: first we will review the Dirac quantization for the RS field. Then, we will apply it to ${\cal L}_{NEK}$, reproducing
the classical result by Hagen \cite{Hagen}, which has been obtained with the Action Principle. Finally, we will apply the same scheme to ${\L}_P$ and will
show that the same positivity issues arise. We then briefly draw our conclusions.

\section{DIRAC QUANTIZATION FOR THE RS FIELD}

We will perform quantization via the Dirac bracket formalism generalized to include fermions (see appendix B).
To do so, we will introduce intermediary brackets for the trivial constraints following \cite{BAAKLINI_ETAL} (which used them for
the free theory) and \cite{HASUMI_ETAL} (which used them for the RS coupled to electromagnetic fields). This is
algebraically much easier than
using the Dirac formalism for the whole set of constraints and eases comparison with \cite{PASCA},  where the same procedure is followed.
First, we will reproduce the quantization of ${\cal L}_{NEK}$ performed in \cite{Hagen} with the action principle, and then we will
quantize ${\cal L}_P$.

The general Lagrangian for the interacting RS, scalar and spinor fields (The interaction between the scalar and spinor fields are not of interest in this context) reads :
\br
{\cal L}&=&{\cal L}_{RS}+{\cal L}_\psi+{\cal L}_\phi+ {\cal L}_P ~~{\mbox or}~~ {\cal L}_{NEK},\label{Lag}
\er
where
\br
{\cal L}_{RS} &=& \bar{\Psi}^\mu \Lambda_{\mu \nu} \Psi^\nu ,~~~
{\cal L}_{\psi}=\bar{\psi} (i\slashed{\partial}-m_\psi)\psi,~~~
{\cal L}_{\phi}=1/2(\partial_\mu \phi \partial^\mu \phi-m_\phi^2\phi),\label{RS}
\er
with
\begin{equation}
 \Lambda_{\mu \nu}=-(i\slashed{\partial}-m)g_{\mu \nu} - iA (\partial_\mu \gamma_\nu + \partial_\nu \gamma_\mu)
                        - iB \gamma_\mu \slashed{\partial} \gamma_\nu - mC \gamma_\mu \gamma_\nu,\label{kin}
\end{equation}
where $A \neq -\frac{1}{2}$, $B=\frac{3}{2}A^2+A+\frac{1}{2}$, $C=3A^2+3A+1$.
The structure of constraints is greatly simplified when $A=-1$: in that case $\Lambda_{00}=0$, so ${\cal L}_{RS}$ becomes independent
of $\dot{\Psi}^0$ . The condition that interactions do not reintroduce a
dynamics for $\Psi^0$ constraints
the possible values for $Z$: $\frac{1}{2}$ in ${\cal L}_{NEK}$ and $-\frac{1}{2}$ for ${\cal L}_P$ (see appendix A).
Then we have ($\epsilon^{\mu\nu\lambda\rho}=i/2\{-i\sigma^{\mu\rho},\gamma^\nu \}$)

\begin{eqnarray}
   \Lambda_{\mu\nu} &=& -\epsilon^{\mu\nu\lambda\rho}\gamma_5\gamma_\lambda\partial_\rho +i m\sigma^{\mu\nu} \nn\\
   {\cal L}_P &=& -g\bar{\Psi}_\mu[\Lambda^{\mu \nu}(m=0)\psi]\partial_\nu\phi+hc.,\nn\\
{\cal L}_{NEK} &=& g \bar{\Psi}_\mu i\sigma^{\mu\nu} \psi \partial_\nu \phi + hc.\label{A_1}
\end{eqnarray}
Next we define the momenta $\Pi_{f,f^\dag} = \frac{\partial {\cal L}}{\partial \dot{f},\dot{f^\dag}}$ (see appendix B) and using that $\epsilon^{0ijk}\gamma_5\gamma_k=
\epsilon_{ijk}\gamma_5\gamma_k=\sigma_{ij}\gamma_0$ we get the not null ones
\br
\Pi_{\Psi_0}\equiv\Pi_0&=&0,~~~\Pi_{\Psi_0^\dag}=\Pi_{0^\dag} =0\label{Pi0delta}\\
\Pi_i &=& -\Psi^\dag_k  \sigma_{ki},~~~\Pi_{i^\dag} =0 \label{Pidelta}\\
\Pi_\psi &=& i\psi^\dag  + \bin {0} {g\Pi_j (\partial_j \phi)},~~~\Pi_{\psi^\dag}= \bin {0} {g\Pi^\dag_j (\partial_j \phi)}\label{Pinucleon}\\
\Pi_\phi&=&-g\Psi^\dag_i \bin  { \gamma_i \psi} { -\sigma_{ij} (\partial_j \psi)}+\dot{\phi},\label{Pipi}\er
where the upper value in brackets corresponds to the ${\cal L}_{NEK}$ interaction while the lower one to the ${\cal L}_P$ case.

Whenever a degree of freedom $f$ is such that $\dot{f}$ cannot be solved in terms of $f$ and $\Pi_f$, constraints arise. So
\br
\chi_0(x)&\equiv&\Pi_0(x)=0,~~~\chi_{0^\dag}(x)\equiv
\Pi_{0^\dag}(x)=0\label{cPi0delta}\\
\chi_i(x)&\equiv&\Pi_i +\Psi^\dag_k  \sigma_{ki}=0,~~~\chi_{i^\dag}(x)\equiv\Pi_{i^\dag} =0,\label{cPidelta}\\
\chi_\psi&\equiv&\Pi_\psi - i\psi^\dag  - \bin {0} {g\Pi_j (\partial_j \phi)}=0,\label{cPiN1}\\
\chi_{\psi^\dag}&\equiv&\Pi_{\psi^\dag}- \bin {0} {g\Pi^\dag_j (\partial_j \phi)}=0,\label{cPiN2}
\er
are primary constraints. For the RS field $\Psi$  in\rf{cPidelta}  as usually done with the Dirac
field, one can eliminate $\Psi^\dag$ in terms of $\Pi$   directly and
using the identity $\sigma_{ij}\left( \frac{i}{2} \gamma_k \gamma_j \right)
= \delta_{ik}$ we get
\br
\Psi^\dag_i =- \frac{i}{2} \Pi_k \gamma_i \gamma_k. \label{PiPsi}
\er
Then by using the fundamental Poisson brackets given in appendix B we get the nonzero ''intermediary brackets''
\begin{eqnarray}
  \{ \Psi_i(x), \Psi_j^\dagger(y) \}_I &=& \frac{i}{2} \gamma_j \gamma_i \delta^{(3)}(x-y)\nn   \\
  \{ \Psi_0(x), \Pi_0(y) \}_I &=&\delta^{(3)}(x-y)\nn \\
  \{ \phi(x)  , \Pi_\phi(y) \}_I &=& \delta^{(3)}(x-y),\label{intermediate}
\end{eqnarray}
where $\Psi_0,\Pi_0$ are not affected by the eliminated constraint and for the pseudoscalar field
it coincides with the fundamental one since we have not a constraint.
Note that this algebra could at first be achieved via the Dirac procedure(see Appendix B), as done in \cite{HASUMI_ETAL} and \cite{PASCA}.
As can be seen form eq.\rf{cPiN1} for the ${\cal L}_{NEK}$ interaction, we could to get $\psi=-i\Pi_\psi$ in analogy with \rf{PiPsi} and using the fundamental brackets to get

\begin{eqnarray}
  \{\psi(x),\psi^\dag(y)\}_I &=& -i\delta^{(3)}(x-y),\label{psipsi}
\end{eqnarray}
but this procedure is not more valid for ${\cal L}_P$, since it  connects $\psi,\Pi_\psi$,and $\Pi$(or $\Psi^\dag$).
Let us then introduce a second set of intermediate bracket eliminating the conjugate momenta of the spinor field by using the Dirac formalism, the
obtained results will be also valid for ${\cal L}_{NEK}$ making $g=0$ since with this value we get the right constraints for this case in eqs.\rf{cPiN1} and \rf{cPiN2}. In order to find the second intermediate bracket we need
\begin{equation}
\{ \chi_\psi(x) , \chi_\psi^\dagger(y) \}_I = -i \delta^{(3)} (x-y),~~~~\{ \chi_\psi(x) , \chi_\psi^\dagger(y) \}^{-1}_I = i \delta^{(3)}(x-y)
\end{equation}
and  the Dirac brackets for the  scalar, spinor and RS fields look like
\begin{eqnarray}
  \{ \Psi_i(x), \Psi_j^\dagger(y) \}_{II} &=& \left( \frac{i}{2} \gamma_j \gamma_i
                        - i g^2 (\partial_i \phi(x)) (\partial_j \phi(y)) \right)  \delta^{(3)}(x-y)   \nn   \\
  \{ \Psi_i (x), \psi^\dagger (y) \}_{II} &=&  -i g (\partial_i \phi(y)) \delta^{(3)} (x-y) \nn        \\
  \{ \psi(x) , \Psi^\dagger_i (y) \}_{II} &=&  i g (\partial_i \phi(x)) \delta^{(3)} (x-y) \nn        \\
  \{ \psi(x)  , \psi^\dagger(y) \}_{II} &=& -i\delta^{(3)}(x-y)\nn                                   \\
  \{ \Psi_0(x), \Pi_0(y) \}_{II} &=&\delta^{(3)}(x-y)\nn \\
  \{ \phi(x)  , \Pi_\phi(y) \}_{II} &=& \delta^{(3)}(x-y),\label{intermediateII}
\end{eqnarray}
for the ${\cal L}_P$ interaction, while making $g=0$ we get those for the ${\cal L}_{NEK}$ case.
Now  the other primary constraints $\chi_0,\chi_{0^\dag}$, cannot be eliminated as above. Then, we impose the  condition to be preserved in time, that is $\theta_0(x) \equiv \{ \Pi_0(x), H\}_{II} =0$  and get  new  nontrivial secondary constraints. The Hamiltonial density reads

\br
{\cal H}(x) &=& \Pi_\mu(x) \dot{\Psi}^\mu(x) + \Pi_\phi(x) \dot{\phi}(x)+
\Pi_\psi(x) \dot{\psi}(x)+ \mbox{hc.} -{\cal L}(x)\nn\\
&=& -\Psi^\dag_i \epsilon_{ijk}  \gamma_5\partial_k \Psi_j  +\Psi^\dag_i  \bin{ ig\sigma_{ij}\gamma_0\psi\partial_j\phi}{-g\epsilon_{ijk}\gamma_5(\partial_j\psi)(\partial_k \phi)}+hc.
                                      - im \bar{\Psi}_i \sigma_{ij} \Psi_j \nn\\   &+& i \bar{\psi}\mbna\cdot\mbga\psi   +\bar{\psi} m_\psi +1/2\mbna\phi\cdot\mbna\phi                                    +1/2m_\phi^2\phi^2\nn\\
   &+&\Psi^\dag_0 \left[ (\sigma_{ji}\partial_j  - m \gamma_i)\Psi_i + \bin {-g \gamma_i \psi \partial_i \phi} { -g\sigma_{ji} (\partial_j \psi)(\partial_i \phi)} \right] + hc. \label{ham}
\er
and thus we get
\begin{eqnarray}
  \theta_0 &=& \{ \Pi_0, H \}_{II} = (\partial_j \Psi_i^\dag\sigma_{ji}  + m \Psi_i^\dag\gamma_i) + \bin {g\psi^\dag \gamma_i\partial_i \phi} { -g(\partial_j \psi^\dag)(\partial_i \phi)\sigma_{ji} }       \nn \\
   \theta_{0^\dagger}&=& \{ \Pi_{0^\dag}, H \}_{II}=(\sigma_{ji}\partial_j  - m \gamma_i)\Psi_i + \bin {-g \gamma_i \psi \partial_i \phi} { -g\sigma_{ji} (\partial_j \psi)(\partial_i \phi)}.\label{secondary1}
\end{eqnarray}

Observe that $\theta_0$ and $\theta_{0^\dagger}$ are the coefficients in \rf{ham} of $\Psi_0$ and $\Psi_{0^\dagger}$ respectively,
so the latter are indeed Lagrange multipliers. Observe that
$\{ \Pi_0 , \theta_0 \}_{II} = \{ \Pi_{0^\dagger} , \theta_0 \}_{II} = \{ \Pi_0 , \theta_{0^\dagger} \}_{II} =
\{ \Pi_{0^\dagger} , \theta_{0^\dagger} \}_{II} = 0$
but for $m\neq 0$  (as will be shown) $\{ \theta_0, \theta_{0^\dagger} \}_{II}\neq 0$.
By imposing $\{ \theta_0, H\}_{II} =0$ and $\{ \theta_{0^\dagger}, H\}_{II} =0$ in order to preserve $\theta_0,\theta_{0^\dagger}$ in time,
we get  tertiary constraints proportional to $\Psi_0$ and $\Psi^\dagger_0$  giving nonzero brackets with
$\Pi_0=0$ and $\Pi_0^\dagger=0$. Nevertheless the only effect
of those tertiary constraints is to determine $\Psi_0$ and $\Psi^\dagger_0$, which are no dynamical
as was seen above, so the relevant Dirac algebra reduces to $\theta_0$ and $\theta_{0^\dagger}$.
Consequently, the only nontrivial bracket to consider in the Dirac procedure is $\{ \theta(x)_0,\theta_{0^\dagger}(y) \}_{II}$, which value as can be obtained from eq.\rf{secondary1} and depends on the NEK or P case.

\subsection{Conventional coupling}

For the ${\cal L}_{NEK}$ conventional coupling the second intermediate brackets are obtained from
\rf{intermediateII} making $g=0$ and read

\begin{eqnarray}
  \{ \Psi_i(x), \Psi_j^\dagger(y) \}_{II} &=& \frac{i}{2} \gamma_j \gamma_i \delta^{(3)}(x-y)\nn   \\
  \{ \Psi_0(x), \Pi_0(y) \}_{II} &=&\delta^{(3)}(x-y)\nn \\
  \{ \psi(x)  , \psi^\dagger(y) \}_{II} &=& -i\delta^{(3)}(x-y)\nn                                   \\
  \{ \phi(x)  , \Pi_\phi(y) \}_{II} &=& \delta^{(3)}(x-y),\label{NEKintermediateII}
\end{eqnarray}
from which we can get using \rf{secondary1}
\begin{equation}
\{ \theta_0(x), \theta_{0^\dagger}(y) \}_{II} = \frac{3i}{2} \delta^3(x-y)
                                                         \left(m^2 - \frac{2g^2}{3}({\bf \nabla}\phi)^2 \right).\label{secondary2}
\end{equation}
Observe that the quantity between parenthesis in the rhs, which is reported in \cite{Hagen} and \cite{PASCA}, can become zero at the classical
level for certain values of the gradient.
In   eq.(30) of ref. \cite{PASCA} it appears in the path integral where the theory is supposed to be quantized,
which is finally not developed.
Here only it is stated that a kind of noncovariance in the measure (where the expressions have several missprints, see Appendix C) ussually
happens and may be cancelled, and raises the question (without answering it) if such cancellation verifies or not in this case.
The only condition asked to quantize the theory in \cite{PASCA}is that $R(x)\equiv 3/2 i (m^2 - \frac{2g^2}{3}g^2({\bf \nabla}\phi(x))^2)  \neq 0$,
to avoid a violation of DOF counting.
But the problem is more
serious: as shown in \cite{JS} and \cite{Hagen}, the actual problem is that $R$ flips sign, making the Hilbert space
non positive-definite. So, the theory was not actually quantized in ref. \cite{PASCA}, the signature problem means that
it is not possible \cite{JS}, \cite{Hagen}.
To show the existence of this signature problem
we will construct the simplest nontrivial anticommutator, the one between of spinor fields, and evaluate it between states on a classical
scalar background.
The Dirac brackets between $\psi$ fields that take into account  the secondary constraints are, by using eqs.\rf{secondary1} and \rf{NEKintermediateII},
\begin{eqnarray}
\{ \psi(x), \psi^\dagger(y) \}_D &=&   \{ \psi(x), \psi^\dagger \}_{II} - \\ \nonumber
             && \int d^3z \; d^3z'  \{\psi(x),\theta(z)\}_{II} \left( \{\theta_0(z),\theta_{0^\dagger}(z')\}_{II} \right)^{-1} \{\theta^\dagger(z'),\psi^\dagger(y)\}_{II}\nn\\
             &=&-i\frac{ \delta^3(x-y)}{1-\frac{2g^2}{3m^2}(\nabla\phi)^2} ,\label{DiracN}
\end{eqnarray}
while the corresponding quantum anticommutator will be
\begin{equation}
[\psi(x) , \psi^\dagger(y)]_+ = \hbar\frac{ \delta^3(x-y)}{1-\frac{2g^2}{3m^2}(\nabla\phi)^2}.
\end{equation}
We will see that the study of the sign definiteness of the space of states, required for consistency, reduces to the possibility of
$R(x)$ vanish when evaluated for quantum states.

Now, let $\ket{f}$ be a coherent state for the $\phi$ field, such that $(\nabla\phi(x)) \ket{f} = (\nabla f(x)) \ket{f}$, being $f(x)$ a
c-number function. The calculation of norms of one-particle spinor states on such background will lead to consider the
quantity
\begin{equation}
\bra{f} [\psi(x) , \psi^\dagger(y)]_+ \ket{f} = \hbar \frac{ \delta^3(x-y)}{1-\frac{2g^2}{3m^2}(\nabla f(x))^2} \bra{f} \ket{f}
\end{equation}
which is not positive definite. Observe that if the gradient is nonzero there is always a reference frame where the norm will flip sign,
this conclusion also was arrived at in ref.\cite{Hagen} from a different prodcedure.
This is the real concern at the quantum level, beyond any consideration of loss of degrees of freedom of the classical
theory in a zero measure region of the configuration space.

\subsection{Spin-3/2 gauge invariant coupling}

We will show, with an argument absolutely paralell to the followed in the previous subsection (which leads to the same results of  \cite{Hagen}) that the
theory is not quantizable for the $\cal{L}_P$ interaction, so it is pointless to develop Feynman rules as in ref. \cite{PASCA}. Of course, in
absence of backgrounds the quantization is the same as the corresponding to the free RS theory, and perturbative expansions of
both ${\cal L}_{NEK}$ and ${\cal L}_P$ are unproblematic.

Using the second intermediate bracket \rf{intermediateII} in \rf{secondary1} we get

\begin{equation}
\{ \theta_0(x),\theta_{0^\dagger}(y) \}_{II} = \frac{3im^2}{2} \left( 1 - \frac{2g^2}{3}({\bf \nabla}\phi)^2  \right) \delta^3(x-y)
\label{finalcondPASCA}
\end{equation}
which, very similarly to the previous case, can vanish for certain values of the gradient at the classical level if $m\neq 0$.
It is interesting to analyze the zero mass limit
since it corresponds to a gauge invariant theory under the transformation $\Psi_\mu \rightarrow \Psi_\mu + \partial_\mu \epsilon$,
where $\epsilon$ is an arbitrary spinor for ${\cal L}_{RS}$. Naively, one should expect (\ref{finalcondPASCA}) to vanish as a consequence
of this gauge invariance thus rendering $\theta_0$ and $\theta_{0^\dagger}$ first class constraints,
in spite there is no general proof that every gauge invariance leads to first class constraints \cite{bookcon}. As a matter
of fact, there are some counterexamples when the gauge symmetry acts trivially \cite{GaugeCounter} and
in this case the gauge symmetry acts trivially on the scalar and spinor fields.
To check it rigorously, observe that(see \rf{A_1} and $S^0\equiv g \Lambda^{0 \nu}(m=0)\psi \partial_\nu\phi$ )
\begin{eqnarray}
Q &=& \int d^3x S^0(x) =-\int d^3x g \gamma_0\sigma_{ji} (\partial_j \psi)(\partial_i \phi)(x),\nn              \\
\end{eqnarray}
and $Q^\dagger$ should be the generators of the gauge symmetry due to the interaction, but its bracket with all fields
vanishes identically (using \rf{intermediateII})
\begin{eqnarray}
\{ Q, \Psi^\dagger_\mu (y) \}_{II} &=& 0 \\
\{ Q, \phi^\dagger (y) \}_{II} &=& 0 \\
\{ Q, \psi^\dagger (y) \}_{II} &=& 0.\nn
\end{eqnarray}
So, although the first terms of $\theta_0$ in eq. \rf{secondary1} $\Lambda^{0 \nu}(m=0)= -\epsilon^{0\nu\lambda\mu}\gamma_5\gamma_\lambda\partial_\mu=- \gamma_0\sigma_{ji}\partial_j$ in the massless limit act as the generator of
$\Psi_\mu \rightarrow \Psi_\mu + \partial_\mu \epsilon$, the last term in spite of being nonzero, does not
generate any gauge transformation. By the same token, the scalar and spinor fields are neutral under this gauge transformation,
which suggests that in a consistent gauge theory should decouple
from the RS field. Observe that $\{Q,Q^\dagger\}=0$ in spite of $\{S^0,S^{0\dag}\}\neq 0$.

To check the signature of the Hilbert space let us proceed analogously as in our previoues subsection for ${\cal L}_{NEK}$. We intend firstly
calculate $\{ \psi(x), \psi^\dagger(y) \}_D$ as in eq.\rf{DiracN} but using now the eq.\rf{finalcondPASCA}
and the commutators of eqs.\rf{intermediateII}. We get
\br
\{ \psi(x), \psi^\dagger(y) \}_D &=& -i\delta^3(x-y),\label{DiracNP}
\er
that is unaffected by the secondary constraint and then, we must pursue looking for possible problems for
\begin{eqnarray}
\{ \Psi_i(x), \Psi^\dagger_j(y) \}_D &=&   \{ \Psi_i(x), \Psi^\dagger_j(y)\}_{II}\nonumber\\
             &-& \int d^3z  d^3z'  \{\Psi_i(x),\theta_0(z)\}_{II}  \{\theta_0(z),\theta_{0^\dagger}(z')\}_{II}^{-1} \{\theta_{0^\dagger}(z'),\Psi_j^\dagger(y)\}_{II}.\nn\\ \label{DiracDP}
\end{eqnarray}
Now by using the corresponding entry for the ${\cal L}_P$ case in eq.\rf{secondary1}, the corresponding brackets \rf{intermediateII}
and eq.\rf{finalcondPASCA}we get

\begin{eqnarray}
\{ \Psi_i(x), \Psi^\dagger_j(y) \}_D &=&   \left[ \frac{i}{2} \gamma_j \gamma_i
                        - i g^2 (\partial_i \phi(x)) (\partial_j \phi(y)) \right]  \delta^{(3)}(x-y)\nonumber\\
             &+& \int d^3z    \{\partial_i^x -i m g^2(\partial_i\phi(x))(\partial_k\phi(z))\gamma_k+{i\over 2} m\gamma_i\}\delta^{(3)}(x-z)\nn\\  &\times&
             \frac{\frac{-2i}{3m^2}}{ \left( 1 - \frac{2g^2}{3}({\bf \nabla}\phi(z))^2  \right)}\nn\\
 &\times&\{\partial_j^{y} -i m g^2(\partial_l\phi(z))\gamma_l(\partial_j\phi(y))+{i\over 2}m\gamma_j\}\delta^{(3)}(z-y),\nn\\ \label{DiracDP}
\end{eqnarray}
where the integral in z' was absorbed by the $\delta(z-z')$ in eq.\rf{finalcondPASCA} and where the property $\partial^z_i \delta^3(z-x)=-\partial^x_i\delta^3(z-x)$ was used.  We can arrange the Dirac bracket as
\begin{eqnarray}
\{ \Psi_i(x), \Psi^\dagger_j(y) \}_D &=&   \left[ \frac{i}{2} \gamma_j \gamma_i
                        - i g^2 (\partial_i \phi(x)) (\partial_j \phi(y)) \right]  \delta^{(3)}(x-y)\nonumber\\
          &+&,\{\partial_i^x -i m g^2(\partial_i\phi(x))(\partial_k\phi(x))\gamma_k+{i\over 2} m\gamma_i\}\nn\\  &\times&
             \frac{\frac{-2i}{3m^2}}{ \left( 1 - \frac{2g^2}{3}({\bf \nabla}\phi(x))^2  \right)}\nn\\
 &\times&\{-\partial_j^{x} -i m g^2(\partial_l\phi(x))\gamma_l(\partial_j\phi(x))+{i\over 2}m\gamma_j\}\delta^{(3)}(x-y),\nn\\ \label{DiracDP1}
\end{eqnarray}
note that in the case of free RS fields ($g=0$) our result conicides with that in ref.\cite{HASUMI_ETAL}($e=0$).

The difficulty in analyzing the signature for the RS states is that the constraint $\theta_{0}$ for the P-case should be enforced (recall that even
the free theory includes negative norm states, which are eliminated only when the constraints are imposed). Let us create a single
RS particle at rest ($\vec{p}=0$) in presence of a scalar background of constant gradient: for definiteness and simplicity let us impose $f(x)=Ax_1$, being $\ket{(A,0,0)}$  a coherent state such that
$\nabla \phi \ket{(A,0,0)} = (A,0,0) \ket{(A,0,0)}$, and absence of any Dirac field quanta. We then built the state $\alpha_i^\dagger \Phi_i \ket{(A,0,0)}$ where $\alpha_i$
is  a vector-spinor coefficient as those appearing in the second quantization expansion of $\Psi_i(x)$, being $\Phi_i = \int d^3 x e^{\xb\cdot\bf0} \Psi_i (x)$ are creation operators of RS quanta at rest, where
to achieve normalization, a regulator volume $V$ should be used. When $\vec{p}=0$ and in absence of nucleon quanta the constraint $\theta_{0}$
implies then $\gamma_i \alpha_i = 0$. One such state is $\vec{\alpha} = (\gamma_2 \chi, \gamma_1 \chi, 0)$ for some constant nonzero
spinor $\chi$, where on time  $\vec{\alpha}^\dagger = (-\chi^\dagger \gamma_2, -\chi^\dagger \gamma_1, 0)$. Let us calculate the norm
\begin{equation}
  \bra{(A,0,0)} (\Phi_i^\dagger \alpha_i) (\alpha_j^\dagger \Phi_j) \ket{(A,0,0)}=
  \alpha_j^\dagger \bra{(A,0,0)} i \{ \Phi_j, \Phi_i^\dagger  \}_D \ket{(A,0,0)} \alpha_i
\end{equation}
where $\{ \Phi_i, \Phi_j^\dagger  \}_D = \int d^3 x d^3 y \{ \Psi_i(x), \Psi_j^\dagger(y) \}_D$. Since once all field operators
act on the states no $x$ or $y$ dependence results in the rhs of \rf{DiracDP1} except for the dirac Delta from the bracket, the integration over $x$ and $y$
of the delta results in a factor $V$ (the regulated volume) which we absorb in the normalization of $\chi$. We thus get
\begin{equation}
  \bra{(A,0,0)} (\Phi_i^\dagger \alpha_i) (\alpha_j^\dagger \Phi_j) \ket{(A,0,0)} =
  2 \left( 1 + \frac{\frac{1}{2}g^2 A^2}{1-\frac{2}{3}g^2 A^2} \right) \chi^\dagger \chi \bra{(A,0,0)} \ket{(A,0,0)}
\end{equation}
which clearly becomes negative for $A$ large enough.

Before ending, let us remark that the quantization procedure in ref. \cite{PASCA} is flawed, since the invoked decoupling of
the introduced auxiliary fields does not verify (see Appendix C). So, the path integration ends at a similar point as in
eq.(30) of ref. \cite{PASCA}. Such expression is useless unless it is developed to show what happen with the signature.

\section{Concluding remarks}

We have shown that the so called spin 3/2-gauge-invariant coupling ${\cal L}_P$ to the RS field presents inconsistencies
analogous to the ones found by Johnson and Sudarshan \cite{JS}
and by Hagen \cite{Hagen} with the usual $\pi$-derivative ${\cal L}_{NEK}$ interaction.
This proves that consistency conjecture of ${\cal L}_P$ stated in \cite{PASCA} is incorrect.
Observe that the main argument in \cite{PASCA} to claim consistency of ${\cal L}_P+{\cal L}_{RS}$ is gauge invariance
under $\Psi_\mu \rightarrow \Psi_\mu + \partial_\mu \epsilon$ in the massless limit. Nevertheless we have shown
that the gauge invariance of this interaction is trivial, in the sense that the gauge transformation does not transform
the scalar and spinor field (observe that the ``current'' $S^\mu_P$ is conserved identically, without imposing the equations
of motion on the scalar and spinor fields). The mass term breaks this invariance anyway. Our treatment cannot be taken to
the massless limit: the Dirac bracket contains the mass as divissor and our proof of signature problems uses the rest frame
which makes sense only in the massive case, but possibly inconsistencies will arise even at the massless limit since the 
gauge invariant interaction is not trivial in spite of the scalar and Dirac fields being neutral under the gauge invariance.

Our result has a great relevance in the hadron fenomenology community, since often the consistency issue is
invoked in evaluations of work done with ${\cal L}_{NEK}$. Recall that ${\cal L}_{NEK}$ and ${\cal L}_P$
are used to interpret accelerator data, estimate parameters for resonances and other critical tasks in phenomenology.
The present work shows that there is no
basis to dismiss work done with interaction ${\cal L}_{NEK}$ or prefer the use of ${\cal L}_P$ in hadron phenomenology
on the basis of their (in)consistency.

On the other hand, the decades old problem of finding consistent interactions
for spin 3/2 fields still remains.

%
%
  \section*{APPENDIX A: RESTRICTIONS FOR $Z$}

Note that in Eq.\rf{Lag} the interaction  can be expressed as ${\cal L}_{NEK,P}=\bar{\Psi}_\mu S^\mu + hc.$ and
let us discuss about the structure of the $S^\mu$ . Observe that in the free RS lagrangian in \rf{RS}, if $A=-1$ (see Eq.\rf{kin}), there is no term
containing $\dot{\Psi}^0$. So, the equation of motion for $\Psi^0$ is a true constraint, and $\Psi^0$ has no dynamics. It is necessary that interactions
do not change that, or there will be no projection of degrees of freedom, and so no hope to get rid from the unwanted negative-norm sector.
The contribution from interactions to such equations of motion will come from $S^0$. The condition that no term containing
$\dot{\Psi}^0$ will arise is that $S^0$ contains no time derivative of any of the other fields of the theory.
Indeed, suppose that $S^0(\dot{\phi},...)$ for some field $\chi=\psi,\phi$, and consider its equation of motion
(${\cal L}={\cal L}_{RS}+{\cal L}_N+{\cal L}_\pi+\bar{\Psi}^\mu S^{NEK,P}_\mu + \bar{S}^{NEK,P}_\mu \Psi^\mu$):
\br
\frac{\partial {\cal L}}{\partial \chi} - \frac{d}{dt} \frac{\partial {\cal L}}{\partial \dot{\chi}}
&=& \frac{\partial {\cal L}}{\partial \chi} -
\frac{d}{dt} \left( \frac{\partial {\cal L}_{RS}+{\cal L}_N+{\cal L}_\pi}{\partial \dot{\chi}}\right)-
\frac{d}{dt} \left( \frac{\partial S^0}{\partial \dot{\chi}} \bar{\Psi}^0 -  \frac{\partial S^i}
{\partial \dot{\chi}} \bar{\Psi}^i\right)\nn\\
&=&\frac{\partial {\cal L}}{\partial \chi}- \frac{d}{dt} \left( \frac{\partial {\cal L}_N+{\cal L}_\pi}{\partial
\dot{\chi}} \right)-\frac{d}{dt} \left( \frac{\partial S^0}{\partial \dot{\chi}} \bar{\Psi}^0 -  \frac{\partial S^i}
{\partial \dot{\chi}} \bar{\Psi}^i\right),
\label{keeplagrange}
\er
since $\frac{\partial {\cal L}_{RS}}{\partial \dot{\chi}}=0$ for all fields.
It is apparent that if $\frac{\partial S^0}{\partial \dot{\chi}} \neq 0$, a contribution proportional to $\dot{\Psi}^0$ will arise.
This condition fixes the off-shell parameter for ${\cal L}_{NEK}$ (in coincidence with determinations by other means in \cite{Nath}
and \cite{PASCA}) as well as ${\cal L}_P$ (only after this value is fixed ${\cal L}_P$ coincides to the interaction proposed
in \cite{PASCA}). Indeed, for ${\cal L}_{NEK}$ ($A=-1$):

\br
{\cal L}_{NEK} &=&  \bar{\Psi}_\nu g(g^{\nu \mu} - (Z+1/2)\gamma^\nu \gamma^\mu) \psi (\partial_\mu \phi)
+ g \bar{\psi} (\partial_\mu \phi)(g^{\mu \nu} - (Z+1/2)\gamma^\mu \gamma^\nu) \Psi_\nu \nn
\er
and so:

\begin{equation}
S^0_{NEK} = g (\frac{1}{2}-Z)\dot{\phi}\psi - g (\frac{1}{2}+Z) \gamma^0\gamma^i \partial_i \phi \psi
\end{equation}
so we get $Z=\frac{1}{2}$, and thus
\begin{equation}
S^0_{NEK} = - g   \gamma^i \gamma^0 \psi(\partial_i \phi)
\label{chinath}
\end{equation}
\begin{equation}
S^{\dagger 0}_{NEK} = g (\partial_i \phi^\dag) \bar{\psi} \gamma^i  .
\label{barchinath}
\end{equation}
${\cal L}_P$ can be written for $A=-1$ as
\br
{\cal L}_P &=& g \epsilon^{ \rho  \alpha \beta \nu }
            \left[
              \bar{\Psi}^\mu (g_{\mu\rho} -(1/2+Z)\gamma_\mu\gamma_\rho)\gamma_5 \gamma_\alpha (\partial_\nu \phi)(\partial_\beta \psi) ]\right.
              \nn\\&+&\left.
               (\partial_\beta \bar{\psi}) (\partial_\nu \phi^\dag)\gamma_5 \gamma_\alpha
               (g_{\mu\rho} -(1/2+Z)\gamma_\mu\gamma_\rho)\Psi^\mu\right],
\er
now should be $Z=-\frac{1}{2}$ to avoid time derivative of fields coupled to $\Psi_0$ then

\begin{equation}
{\cal L}_P = g \epsilon^{\mu \nu \alpha \beta}             \left(
              \bar{\Psi}_\nu \gamma_5 \gamma_\alpha (\partial_\mu \psi)(\partial_\beta \phi)
  + (\partial_\mu \bar{\psi}) (\partial_\beta \phi^\dag)
 \gamma_5 \gamma_\alpha \Psi_\nu
            \right)
\end{equation}
so we get
\begin{eqnarray}
  S^0_P  &=&  g (\partial_i \bar{\psi})(\partial_k \phi) \epsilon^{i0jk} \gamma_j \gamma_5
\end{eqnarray}
\begin{eqnarray}
 S^{\dagger 0}_P  &=&  g (\partial_n \phi) \epsilon^{l0mn} \gamma^0 \gamma_5 \gamma_m (\partial_l \psi)
\end{eqnarray}

We thus see that indeed, within the interaction ${\cal L}_P$ there is a coupling to the spin 1/2 sector. This coupling is not noticeable
at tree level of pure hadron coupling, but shows up in radiative amplitudes and at loop level, as was shown recently in \cite{Nos_background}.

%
%
  \section*{APPENDIX B: DIRAC BRACKETS FOR FERMIONS}

Fermion fields have not a classical limit. However, in order to cope with constrained fermion systems and path integrals in supergravity and string theory,
in the 70's emerged a description of fermion degrees of freedom in the pseudoclassical limit $\hbar \rightarrow 0$. It has been shown that in this
limit fermions are described by Grassman (anticommuting) variables \cite{Casalbuoni}. A generalized Poisson pseudoclassical bracket is defined
for describing pseudoclassical field systems including both c-number and Grassman fields. These are the fundamental brackets adequate to apply the
Dirac quantization scheme in boson-fermion systems. Let $f$ and $g$ be fields, $\lambda$ a parameter, and $\epsilon(h)=0$ if $h$ is bosonic,
$\epsilon(h)=1$ if it is fermionic. Then the Poisson bracket $\{ , \}$ obeys
\begin{eqnarray}
\{ f , g \} &=& (-1)^{\epsilon(f)  \epsilon(g) + 1} \{ g , f \}                           \label{sym} \\
\{ f + h, g \} &=& \{ f , g \} + \{ h , g \}                                              \label{lin} \\
\{ f , \{ g , h \} \} &=& f \{ g , h \} + (-1)^{\epsilon(f) \epsilon(g)}\{ f , h \} g     \label{Lei} \\
\{ f , \lambda \} &=& 0                                                                   \label{scalar}
\end{eqnarray}
Then, for each field $f,f^\dagger$ in the theory, we define its canonical momentum as
\begin{equation}
\Pi_f = \frac{\partial {\cal L}}{\partial \dot{f}},\ \ \Pi_{f^\dagger} = \frac{\partial {\cal L}}{\partial \dot{f^\dagger}}
\label{canonical}
\end{equation}
with the caution that, if $f$ is fermionic, the derivative with respect to it is also anticommuting, so it is important to distinguish
between left or
right derivation. We will adopt right derivation, that is, the derivative of ${\cal C}q$ with respect to the
Grassman variable $q$ will be always ${\cal C}$, but
the derivative of $q{\cal C}$ will be $-1$ if ${\cal C}$ is a Grassman number, and $1$ if it is a c-number.
Observe that for bilinear Lagrangians like ${\cal L}_{RS}$, if all derivatives act on the fields on the right (as it is our case,
\cite{BAAKLINI_ETAL}
and \cite{EndoKimura}, canonical momenta are the same one should obtain from treating $\Psi^\mu$ as c-numbers, but in the case
of symmetrized lagrangians
like in \cite{PASCA} and \cite{HASUMI_ETAL} one must be careful in considering the anticommutation between the Grassman
derivative and $\Psi^\mu$ in the terms
in which derivatives act on $\bar{\Psi}^\mu$. Observe also that if one defines canonical momenta as left derivatives instead
of right, the RS lagrangian
would be minus our ${\cal L}_{RS}$ in order to yield a positive definite spectrum.

We define the ''Poisson brackets'' fulfilling  the properties \rf{sym}-\rf{scalar}
as
\br
\left\{ f(x),g(y) \right\} &=&
\int d^3z \left[ {\partial f(x)\over \partial \Psi(z)}
{\partial g(y)\over \partial\Pi(z)} + (-1)^{\epsilon(f)  \epsilon(g) + 1} f \leftrightarrow g\right.
\nn\\
&+&\left.
{\partial f(x) \over \partial \Psi^\dagger(z)}
{\partial g(y) \over \partial \Pi^\dagger(z)} + (-1)^{\epsilon(f)  \epsilon(g) + 1}f \leftrightarrow g
\right],\label{A0}
\er
from which the so called ''fundamental (equal time) Poisson breackets'' are obtained:
\begin{equation}
\{ f(x) , \Pi_f(y) \} = \delta^3(x-y).
\label{FP}
\end{equation}
These, are not always compatible with \rf{canonical} since $f$ and $\Pi_f$ are not in general independent variables appearing  constraints $\Omega_k = 0$ relating them.

To construct a consistent Poisson algebra, the Dirac procedure is used: the fundamental Poisson brackets between constraints are calculated using \rf{FP}
\begin{equation}
C_{k l}(x,y) = \{ \Omega_k(x) , \Omega_l(y) \}\label{matconst},
\end{equation}
and the linearity property \rf{lin} is used. So we get the new ``Dirac'' brackets as follows:

\br
\{ f(x) , g(y) \}_D &=& \{f,g\}\nn\\&-&(-1)^{\epsilon(f)  \epsilon(g) + 1}\int d^3z \; d^3z' \{ f(x) , \Omega_k(z) \} \; C^{-1}(z,z')_{k l} \; \{ \Omega_l(z'), g(y) \}.\nn\\\label{Diracbra}
\er
These Dirac brackets are also Poisson brackets in the sense that they obey \rf{sym}-\rf{scalar}, but they are now consistent
with the constraints.

For the RS case the  Poisson Bracket between  a field
$f,g =\Psi_{\mu,m}, \Pi_{\mu,m}=\partial{\cal L}/ \partial(\dot{\Psi}^\mu_m)$,
 or $\Psi^\dagger_{\mu,m},\Pi_{\Psi^\dagger \mu,m}=
\partial{\cal L}/ \partial(\dot{\Psi^\dagger})^\mu_m$ is

\br
\left\{ f(x)_{\mu,m},g(y)^\nu_n \right\} &=&\nn\\
&&\hspace{-1cm}\int d^3z \left[ {\partial f(x)_{\mu,m} \over \partial \Psi(z)_{\alpha,a}}
{\partial g(y)^\nu_n \over \partial\Pi(z)^\alpha}_a + f \leftrightarrow g
+
{\partial f(x)_{\mu,m} \over \partial \Psi^\dagger(z)_{\alpha,a}}
{\partial g(y)^\nu_n \over \partial\Pi(z)^\alpha_{\Psi^\dagger a}} + f \leftrightarrow g
\right],\nn\\\label{A1}
\er
where a,m,n=1,2,3,4 are spinor matrix indexes while $\alpha, \mu, \nu$ = 0,1,2,3 are Lorentz ones.
The fundamental equal time Poisson brackets for the RS field are
\br
\left \{ \Psi(x)_{\mu,m},\Pi(y)^\nu_n \right \} &=& g_\mu^\nu \delta_{m,n} \delta^3(x-y)\nn\\
\left \{ \Psi^\dagger(x)_{\mu,m},\Pi^\dagger(y)^\nu_n \right \} &=& g_\mu^\nu \delta_{m,n} \delta^3(x-y)
,\label{A2}
\er
with other combinations of $\Psi,\Pi,\Psi^\dagger,\Pi^\dagger$ vanishing, and where we have used that
$\frac{\partial f(x)}{\partial f(y)} = \delta(x-y)$, while for the Dirac spinor $\psi$ and
the scalar $\phi$ will be
\br
\left \{ \psi(x)_{m},\Pi_\psi(y)_n \right \} &=& \delta_{m,n} \delta^3(x-y)\nn\\
\left \{ \phi(x)_{\mu,m},\Pi_\phi(y)\right \} &=& \delta^3(x-y).\label{A9}
\er

%
%
  \section*{Appendix C}

We will devote this appendix to the path integral treatment of RS interactions in ref. \cite{PASCA}. Let us start with
${\cal L}_{NEK}$. Observe that in expression (29) of \cite{PASCA} the square root in the LHS is not
carried to the RHS, nor to expression (30). Observe also that in the simplification of the
determinant (29) invoked in (30) there should be a factor $R^2$ which is field dependent and thus
cannot be ommitted. The precise form of the determinant is however unimportant for the development of
\cite{PASCA}, since it is not used to derive any result.

On the other hand, the treatment of ${\cal L}_P$, whithout introducing St\"uckelberg parameters
(our treatment), leads
to a determinant very simmilar to that in eq. (29) of \cite{PASCA}, whith
$R=\frac{3im^2}{2} \left( 1 - \frac{2g^2}{3}g^2({\bf \nabla}\phi)^2  \right)$
and $-g\sigma_{ji} (\partial_j )(\partial_i \phi)$
instead of $g\gamma_i \partial_i \phi$ in the matrix positions (3,5) and (5,3),
where $\partial_j$ acts on the Dirac delta. The reduction is even more involved than in the case of ${\cal L}_{NEK}$.
But \cite{PASCA} did not produce such an expression, since they tried another way introducing an auxiliary field $\xi$.
They did so through a reasoning in line with St\"uckelberg formalism, but they could have produced alternatively the expression
(45) by exponentiating part of the measure, in the spirit of the Faddeev and Popov method. Whatever the method employed,
the result is that
the problematic constraint $\Pi_0=0$ together with $\theta_4=0$
are traded
into first class constraints
(thus dissapearing from the determinant in the Dirac algebra), at the price of adding the St\"uckelberg field $\xi$. But
in passing from expression (45) to (46) it is stated without proof that, due to the gauge invariance, $\xi$ decouples. Observe that
the mentioned gauge invariance is no longer the corresponding to the massless case, but the subtler invariance shown in expression
(A2) of \cite{PASCA}, which is a more restricted kind of gauge transformation since the gauge parameter is no longer arbitrary
but obeys a nontrivial ecuation of motion \cite{Stuec}. Thus, there is no reason to think that the functional integration over
$\xi$ leads to its decoupling nor to
the $\delta$ functionals which lead to the reported Feynman rules. As a mater of fact, the statement that gauge invariance (A2)
implies the decoupling of $\xi$
makes no sense: it is the coupling to $\xi$ what compensates the symmetry breaking of the mass term making the theory gauge invariant
\cite{Stuec}. If $\xi$ is integrated out properly, obviously they must have had arrived to the complicated measure mentioned at the begining
of this paragraph, since the original theory should be recovered.

To see how absurd this result is, observe that we could take the limit ${\cal L}_P \rightarrow 0$, so
(46) would imply that the free RS is equivalent
to a theory without the constraints $\Pi_0=0$ and $\theta_4=0$. This is in contradiction with the developement of section II of
\cite{PASCA} itself.
This mistake carries over to the Feynman rules, as stated above.
The right procedure would be, if one is to work with
first class constraints instead of second class ones, to retain the field $\xi$, whith their own Feynman rules, and then
make the gauge choice at the level of the Hilbert space.
In fact, other authors have done generalizations of the St\"uckelberg formalism to RS fields in the past \cite{Watanabe} and
found it to be quite tricky, requiring for instance the introduction of two spin 1/2 St\"uckelberg fields instead
of one.

So, once this flaw in the quantization procedure is pointed out, we get to a situation simmilar to that after expression (30)
of \cite{PASCA}: a functional integration with a complicated measure with wich it is very difficult to proceed with
path integrals. Since afterwards we have shown that the theory is not quantizable (since the Hilbert space is not positive definite
nor semidefinite) there is no point in trying to get any Feynman rules or quantization, by any means.

\section{acknowledgement}

\end{document}